\documentclass[twocolumn,showpacs,superscriptaddress,prl]{revtex4}

\usepackage{graphicx}
\usepackage{subfigure}
\usepackage{wasysym}
\usepackage{color}
\usepackage{esvect}
\usepackage{textcomp}
\usepackage[dvipdfm]{hyperref}
\usepackage{epsfig}

\begin{document}

\title{Giant spin Hall conductivity in platinum at room temperature}

\author{Chee Weng Koong}\email{Koong\_Chee\_Weng@dsi.a-star.edu.sg}
\affiliation{National University of Singapore Graduate School for Integrative Sciences and Engineering, Singapore 117456, Singapore}
\affiliation{Institute of Materials Research \& Engineering, Singapore 117602, Singapore}

\author{Berthold-Georg Englert}
\affiliation{Department of Physics, National University of Singapore, Singapore 117542, Singapore}
\affiliation{Centre for Quantum Technologies, National University of Singapore, Singapore 117543, Singapore}

\author{Christian Miniatura}
\affiliation{Institut Non Lin\'{e}aire de Nice, UNS, CNRS; 1361 route des Lucioles, F-06560 Valbonne, France}
\affiliation{Centre for Quantum Technologies, National University of Singapore, Singapore 117543, Singapore}
\affiliation{Department of Physics, National University of Singapore, Singapore 117542, Singapore}

\author{N. Chandrasekhar}
\affiliation{Institute of Materials Research \& Engineering, Singapore 117602, Singapore}

\date{\today}

\begin{abstract}
We have demonstrated the electrical generation and detection of spin polarization by the spin Hall effect (SHE) in platinum. The experiment was performed in a non-local geometry without the use of ferromagnetic materials or magnetic field. We designed a circuit that uses the SHE to convert a charge current to a spin current, and the inverse SHE to convert the spin current back into a charge signal. The experiments were carried out for temperatures from 10 K up to 290 K. We extracted the spin Hall conductivity and spin diffusion length from the data with the aid of a spin diffusion model, and found the values of 1.1 $\times 10^6$ $\Omega^{-1}$m$^{-1}$ and 80~nm, respectively, at 290 K. The spin Hall conductivity is two orders of magnitudes larger than the previous record of $3.3\times10^4$ $\Omega^{-1}$m$^{-1}$. This observation may have many potential applications in spintronics devices.
\end{abstract}

\maketitle

The spin Hall effect (SHE) refers to the non-equilibrium spin accumulation near sample boundaries when an electrical current is applied. The spin accumulation is a result of the spin-orbit induced spin separation of electrons in the direction transverse to the applied current. The effect was first proposed by D'yakonov and Perel' in 1971 and rediscovered by Hirsch in 1999 \cite{SHE}. From a technical point of view, SHE is noteworthy because it can be used to generate spin currents with 100\% spin polarization. In addition, its reciprocal effect, the inverse SHE, can be used to detect spin-polarized currents. The unique ability to generate and detect spin polarization in a non-ferromagnetic material gives rise to new possibilities for integrating this technology with existing or novel spintronics devices \cite{patent:20090161265,Behin2010}.

Experimental studies and theoretical calculations have shown that SHE in metals are robust against disorder and have larger spin Hall conductivity (SHC) than their semiconductor counterparts \cite{FMR,SpinInjection1,SpinInjection2, PhysRevLett.95.156601,guo:096401}. Early studies of the SHE in Au, Al and Pt have revealed that the SHC in Pt is significantly larger than in these other metals \cite{Koong2007}. Subsequent theoretical studies \cite{guo:096401} and experimental results \cite{SpinInjection1} provided corroborating evidence that Pt has one of the largest SHC among the metals at room temperature. This distinctive feature makes Pt a promising candidate for spintronics applications.

\begin{figure}
\includegraphics[scale=0.6]{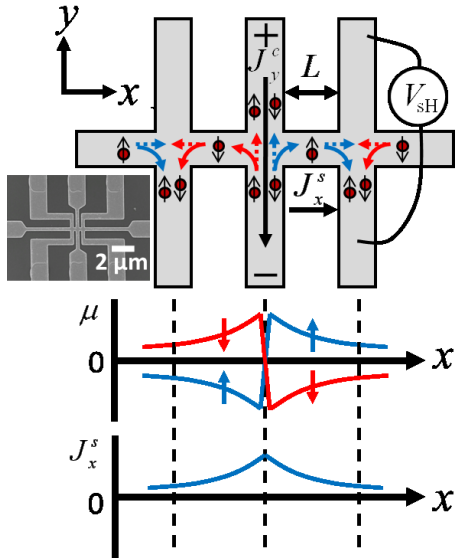}
\caption{(color online) (Top) A charge current $J^\mathrm{c}_y$ flows in the negative $y$-direction, leading to a spin separation in the transverse $x$-direction. The spin imbalance diffuses and, in steady state,  a counter current of opposite spin electrons ensures that only a pure spin current $J^\mathrm{s}_x$ is flowing in the positive $x$-direction. At the next junction, both spin species are deflected to the {\it same} side in the negative $y$-direction, creating a conventional Hall signal across the side arm. The spin Hall voltage $V_\mathrm{sH}$ observed gives rise to an ohmic current in the $y$-direction that compensates for the spin current. The scanning electron micrograph of a typical device is shown at the bottom left corner of the schematic. (Bottom) Spatial dependence of the spin-up and spin-down electrochemical potentials $\mu$ and spin current $J^\mathrm{s}_x$ along the $x$-direction.} \label{Figure1}
\end{figure}

The current experimental techniques for measuring the metallic SHE are current injection from a ferromagnet \cite{SpinInjection1,SpinInjection2} and ferromagnetic resonance-induced spin pumping \cite{FMR}. Both methods have drawbacks because they require ferromagnetic contacts to generate and/or detect spin-polarized electrons. The ferromagnetic contacts induce an inhomogeneous stray field which accelerates the dephasing of the spin \cite{meier:172501}. Furthermore, the ferromagnetic contacts absorb the spin polarization and suppress the spin accumulation in the metal~\cite{kimura:3795}. To understand the underlying SHE mechanism, it is critical to have a proper characterization technique without any spurious influence. We used an all-electrical non-local technique with a ``double H" geometry, which is reported in a earlier publication by the authors \cite{Koong2007}, to characterize the SHE. Our technique does not require any magnetic materials or magnetic fields, and thus offers a radically different way to study the SHE. The ``double H" device is made up of three parallel conductors separated by a distance $L$ and a bridging conductor, as shown in Fig.~\ref{Figure1}. Samples with different $L$, where 80 nm $\leq L \leq$ 450~nm, were sputtered to thickness $t=$ 50 nm concurrently to minimize differences in the thickness and the microscopic structures. The width $w$ of the conductor is 200 nm.

The basic concept behind the design is as follows (See Fig.~\ref{Figure1}). First, a charge current $J^\mathrm{c}_y$ is applied in the middle vertical conductor and a transverse spin current $J^\mathrm{s}_x$ is generated due to the SHE. The transverse current carries electrons away from the middle Hall cross and along the bridging conductor. In steady state, a counter current of opposite spin electrons ensures that there is only a pure spin current $J^\mathrm{s}_x$ and no net electric current flowing in the $x$-direction. As the spin-up and spin-down electrons move in opposite directions, both spin species are now scattered to the {\it same} side of the conductor. At steady state, a charge imbalance builds up that compensate for the spin current. This spin Hall voltage $V_\mathrm{sH}$ is observed in the adjacent Hall arm. This ``double H" geometry is also studied by two other groups \cite{PhysRevB.70.241301, abanin:035304}, and they both concluded that the lateral circuit is suitable for the investigation of the SHE.

\begin{figure}
\includegraphics[scale=0.26]{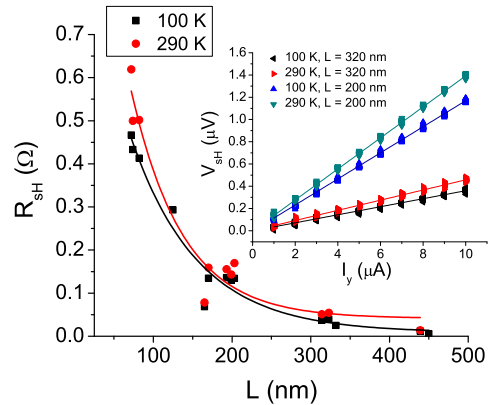}
\caption{(color online) The measurements of spin Hall resistance $R_\mathrm{sH}$ indicate an exponential decay as a function of distance $L$ at 100 K and 290 K. The data are fitted to Eq. (\ref{eq6}) and the spin diffusion
length ${\rm \lambda_\mathrm{s}}$ and spin Hall conductivity ${\sigma_{xy}}$ were determined to be $(91\pm15)$~nm, $(1.5\pm0.3) \times 10^6$~$\Omega ^{-1}$m$^{-1}$ at 100 K and $(80\pm11)$~nm, $(1.1\pm0.2) \times 10^6$~$\Omega ^{-1}$m$^{-1}$ at 290 K. The error bars are in most cases smaller than the symbol, and hence error bars are not represented in the graph. The deviation of the points from the fit is likely to be caused by the microscopic differences among the samples. The representative graph of the $V$-$I$ measurements for $L~=~200$~nm and $L = 320$~nm is shown in the inset. 60 readings were taken for each incremental steps in the applied current $I{_y}$.
} \label{Figure2}
\end{figure}

A three-point current reversal technique, with a synchronized DC current source and nanovoltmeter, is used to eliminate the effect of the thermoelectric voltages from the measurements. The representative $V$-$I$ measurements at 100 K and 290 K, where $L = 200$ nm and $320$~nm, is shown in the inset in Fig.~\ref{Figure2}. One observes that the voltage signal increases linearly with the applied current, which indicates that the Joule heating has no effect on our measurements. We define the spin Hall resistance by $R_\mathrm{sH} =V{_\mathrm{sH}}{/}I{_y}$ where $I_{y}$ is the applied current in the middle vertical conductor. The $R_\mathrm{sH}$ values for all the samples were determined by a linear fit and plotted as a function of the distance $L$, as shown in the main plot in Fig.~\ref{Figure2}.

We used a phenomenological spin diffusion model to describe the SHE and the inverse SHE processes in the devices. When a charge current $J^\mathrm{c}_y={\sigma_{yy}}E_y$  is applied in the middle vertical conductor, the SHE induces a transverse spin current $J^\mathrm{s}_x={\sigma_{xy}}E_y$ in the $x$-direction, where $\sigma_{xy}$ is the spin Hall conductivity and $\sigma_{yy}$ is the Drude conductivity. Upon eliminating $E_y$ from the equations, the spin current at the edge of the Hall cross is expressed as $J^\mathrm{s}_x=({\sigma_{xy}}{/}{\sigma_{yy}})J^\mathrm{c}_y$.

As no electrical field is applied in the $x$-direction, the spin transport can be described by a diffusive transport model. Thus, for $x\geq 0$, the spin current  $J^\mathrm{s}_x(x)=J^\mathrm{s}_x(0)\exp\left(-{x}/{\lambda_\mathrm{s}}\right)$ takes the usual exponential form, where $J^\mathrm{s}_x(0)$ is the spin current at the edge of the first Hall cross and $\lambda_\mathrm{s}$ is the spin diffusion length \cite{JPSJ.77.031009}. At the second Hall arm, where $x=L$, the spin current $J^\mathrm{s}_x(x)$ gives rise to an electrical field $E_y(x)={J^\mathrm{s}_x(x)}{/}{\sigma_{yx}}$ in the $y$-direction due to the inverse SHE. Using the earlier definition of $R_\mathrm{sH}$ and taking the Onsager reciprocal relations ${\sigma_{yx}}={\sigma_{xy}}$ into account, we have
\begin{equation} \label{eq6}
R_\mathrm{sH}(x)=\left( \frac{\sigma_{xy}^2}{t\sigma_{yy}^3}
\right)\exp \left( -\frac {x}{\lambda_\mathrm{s}}\right).
\end{equation}

\noindent Using Eq.~(\ref{eq6}) and a least-square exponential fit of the data in Fig.~\ref{Figure2}, we extracted ${\rm \lambda_\mathrm{s}}$ and ${\sigma_{xy}}$ at 100 K and 290 K, as shown in Table~\ref{Table1}. A standard four-point-probe measurement was used to determine ${\sigma_{yy}}$ at the same temperatures and the values are presented in Table~\ref{Table1}.

\begin{table}
\tabcolsep 5mm \caption{The values of the spin diffusion length $\lambda_\mathrm{s}$, the spin Hall conductivity $\sigma_{xy}$, and the Drude conductivity $\sigma_{yy}$. $\lambda_\mathrm{s}$ and $\sigma_{xy}$ are extracted from the data in Fig.~\ref{Figure2} with a least-square fit to Eq.~(\ref{eq6}). ${\sigma_{yy}}$ is determined by standard four-point-probe measurements.}
\centering
\begin{tabular}{c c c }
\hline \hline
 & 100 K & 290 K\\
 \hline
$\lambda_\mathrm{s}$ (nm)                         & $91\pm15$    & $80\pm11$  \\
$\sigma_{xy}$ ($10^6$ $\Omega ^{-1}$m$^{-1})$     & $1.5\pm0.3$  & $1.1\pm0.2$\\
$\sigma_{yy}$ ($10^6$ $\Omega ^{-1}$m$^{-1}  )$   & $3.6\pm0.3$  & $2.6\pm0.1$\\
    \hline\hline
\end{tabular}\label{Table1}
\end{table}

The value found  for ${\sigma_{xy}}$ at 290 K is two orders of magnitudes larger than the previously reported figures of $10^4$~$\Omega^{-1}$m$^{-1}$ \cite{SpinInjection1}. The huge difference in the two sets of numbers could be related to the presence of ferromagnetic materials in the other experimental setup. Indeed, Pt films are likely to be contaminated during the deposition of the ferromagnets, leading to magnetic impurity disorder in the structure. As mentioned earlier, the ferromagnet stray fields reduce the lifetime of the spin current \cite{meier:172501} and the ferromagnetic contacts also absorb the spin polarization \cite{kimura:3795}. All these effects diminish the value of ${\sigma_{xy}}$. Our value for ${\rm \lambda_\mathrm{s}}$ in Pt is within the experimental range of 14 nm measured at 4.2 K \cite{kurt:4787} in a current-perpendicular-to-plane geometry and 120~nm at 5 K \cite{Han2007} measured with a current-in-plane lateral spin valve.

\begin{figure}
\includegraphics[scale=0.29]{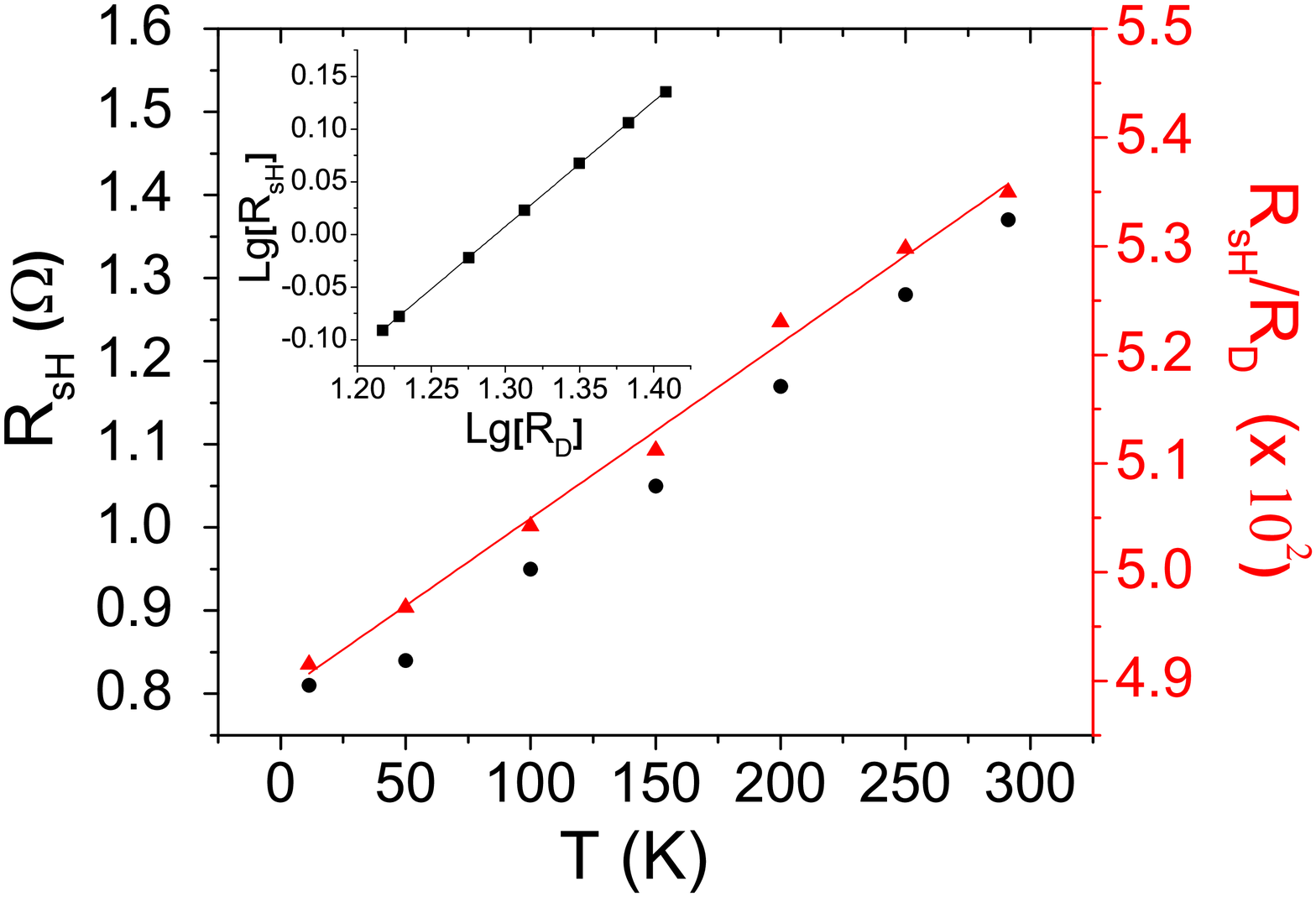}
\caption{(color online)
Experimental results of the spin Hall resistance $R_\mathrm{sH}$ (\textbullet) and the ratio $R_\mathrm{sH}/R_\mathrm{D}$ (\textcolor{red}{{\UParrow}}) as a function of temperature, where $R_\mathrm{D}$ is the Drude resistance and $L$~=~115~nm. The temperature dependence of the ratio is a signature of the SHE. Inset: The slope of $R_\mathrm{sH}$ against $R_\mathrm{D}$ in log-log scale is 1.2, which suggests that the extrinsic skew scattering is the dominant mechanism in Pt. The error bars are in most cases smaller than the symbol, and hence they are not represented in the graphs.}\label{Figure3}
\end{figure}

Beside the SHE, another possible explanation for the observation of the non-local resistance is related to the ``leakage" of current flux from the middle vertical conductor. Such a charge current spreading also leads to an exponential decay of the nonlocal resistance \cite{mihajlovic:166601}. However the corresponding ``leakage" decay constant would be $w{/}\pi =64$ nm in our case, significantly lower than our measured values of $\lambda_s$. The charge current spreading artifact is thus unlikely to explain our Fig.~\ref{Figure2}. Furthermore, if the non-local resistance is ohmic in nature, one would expect the same temperature dependence for both the $R_\mathrm{sH}$ and the Drude resistance $R_\mathrm{D}$. We have measured the temperature dependence of $R_\mathrm{D}$ and $R_\mathrm{sH}$ for temperatures from 10 K up to 290 K and plotted both $R_\mathrm{sH}$ and $R_\mathrm{sH}/R_\mathrm{D}$ in Fig.~\ref{Figure3}. $R_\mathrm{sH}/R_\mathrm{D}$ has a strong temperature dependence which proves that the non-local resistance is not related to the current flux ``leakage".

Currently, the mechanism behind the SHE in Pt is not well understood. However the general consensus is that the SHE mechanism is similar to the anomalous Hall effect (AHE). Onoda {\it et al.} integrated the various intrinsic and extrinsic mechanisms in AHE into a unified theoretical framework \cite{PhysRevLett.97.126602}, where it was shown that the dominant effect is the extrinsic skew scattering in the clean limit. We performed an energy-dispersive X-ray analysis and secondary ion mass analysis and concluded that the impurities concentration in our film is less than 1 ppm. In the extrinsic SHE mechanism, the spin Hall
resistivity is related to the conventional resistivity by $\rho_{xy} = A\rho^n_{xx}$, where $A$ is a constant. If $n= 1$, then the effect is attributed to the skew-scattering mechanism \cite{1958Phy....24...39S} whereas if $n=2$, the effect is attributed to the side-jump mechanism \cite{PhysRevB.2.4559}. We plotted the log-log graph of $R_\mathrm{sH}$ as a function of $R_\mathrm{D}$ in the inset in Fig.~\ref{Figure3}. The slope of the fitted line is 1.2, which suggests that skew-scattering is indeed the dominant source for SHE in Pt.

A recent experiment by Mihajlovi\'{c} {\it et al.}~\cite{mihajlovic:166601}, who studied the non-local resistance in Au based on a similar non-local geometry, reported a negative non-local resistance below 80 K. They concluded that this surprising result is due to the combined effect of a positive contribution $R_\mathrm{c}$ from current flux leakage and a negative contribution $R_\mathrm{b}$ from the ballistic transport of electrons at the side arm, and is not attributed to the positive SHE contribution $R_\mathrm{sH}$. A simple comparison between their Au results and our Pt measurements (with similar geometry at room temperature) revealed that our $R$ is a factor of $10^3$ larger. The large discrepancy in the two numbers indicate that the underlying mechanisms in Au and Pt are likely to be different.

\begin{figure}
\includegraphics[scale=0.26]{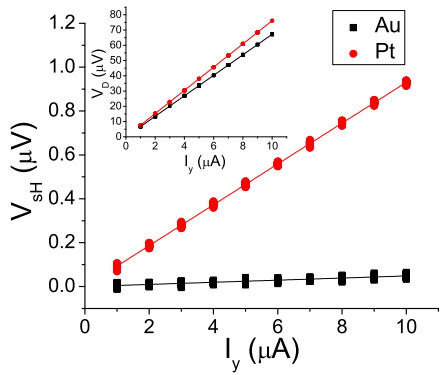}
\caption{(color online) Experimental results for the spin Hall voltage $V_\mathrm{sH}$ against applied current $I_{y}$ for Au and Pt where $L$ = 190 nm at 100 K. The two slopes differ by a factor of about 20, which signals a large spin Hall conductivity ${\sigma_{xy}}$ in Pt. Inset: We infer from the ohmic resistance of Au and Pt measured by conventional four-point-probe measurements that the resistivities of the two films are nearly identical. 60 readings were recorded for each incremental steps in the applied current $I{_y}$ for both graphs.} \label{Figure4}
\end{figure}

We claim that our observation is related to the SHE. We measured one Pt and one Au sample with identical dimensions and temperature to highlight the different mechanisms between the two metals. The non-local resistance $R=R_\mathrm{sH}+R_\mathrm{c}-R_\mathrm{b}$ can be considered to be made up of these three components. We use a counterexample by assuming that $R_\mathrm{sH}$ is negligible or non-existent in Pt. As shown in the main plot of Fig.~\ref{Figure4}, our measurement of $R$ in Pt at 100 K is 20 times larger than in Au. Following the previous equation and putting $R_\mathrm{sH}=0$, this means that either $R_\mathrm{c}$ is significantly larger in Pt, or $R_\mathrm{b}$ is significantly smaller. In the inset of Fig.~\ref{Figure4}, we measured the ohmic voltage and showed that the resistivities of the two films are almost identical. The large difference in the non-local resistance between Au and Pt is not proportional to the difference in their Drude resistance, and thus our observation could not be explained by the current flux ``leakage" $R_\mathrm{c}$. In contrast to the Au experiments in Ref.~\cite{mihajlovic:166601} which reported a negative non-local resistance below 80 K, our non-local resistance measurements for Pt remains positive for temperature as low as 10 K (see Fig.~\ref{Figure3}). Thus we also excluded the ballistic contribution $R_\mathrm{b}$ in Pt as an explanation for our large $R$ values. As neither $R_\mathrm{c}$ nor $R_\mathrm{b}$ could account for our observations, the possible explanation left is that the $R_\mathrm{sH}$ term is the dominating contributor in the observed $R$ values.  We also do not observe any significant spin Hall signal from repeated measurements in Au samples with different $L$. Thus, we conclude that the SHC in Pt is larger than in Au and that our findings are consistent with the conclusions reached in Ref.~\cite{mihajlovic:166601}.

In summary, we have measured the spin Hall conductivity ${\sigma_{xy}}$ in Pt with a ``double H" shaped non-local geometry. Our measurements are purely electrical. The device makes no use of a magnetic field or ferromagnetic materials and thus avoids the complications of spin interaction with stray magnetic field. We found a giant spin Hall conductivity from 10 K up to 290 K --- larger by two orders of magnitudes than previously reported values. Although it is difficult to isolate the spin Hall effect from the other contributions, we showed that the dominant contribution to the measured voltage is the spin Hall resistance $R_\mathrm{sH}$. Most remarkably, the spin Hall effect in Pt is large even at room temperature and remains one of the most promising ways for producing large spin polarization for spintronics applications.

\begin{acknowledgments}
This paper is based upon work supported by the Agency for Science, Technology and Research (A{$^\ast$}STAR) through the Institute of Materials Research \& Engineering (IMRE). The authors would like to thank Gilles Montambaux, Marta Wolak and Mai Van Do for useful discussions. CM acknowledges support by the Centre National de la Recherche Scientifique (CNRS) PICS Grant No.~4159. Institut Non Lin\'{e}aire de Nice is unit\'{e} mixte de recherche UMR 6618 du CNRS. CM and BGE acknowledge support by the France-Singapore Merlion program SpinCold Grant No.~2.02.07. Centre for Quantum Technologies is a Research Centre of Excellence funded by Ministry of Education and National Research Foundation of Singapore.
\end{acknowledgments}

Note added --- After our manuscript was submitted, an independent experimental observation of the SHE in HgTe based on the same ``H" geometry has been reported \cite{Brune2010}. We point out that their results are consistent with our observations in Pt.

\end{document}